\documentclass[%
reprint,
amsmath,amssymb,
aps,
]{revtex4-1}

\usepackage{graphicx}
\usepackage{bm}
\usepackage[mathlines]{lineno}

\begin{document}

\title{Simple particle model for low-density granular flow interacting with ambient fluid}

\author{Hirofumi Niiya}
\email[]{niiya@gs.niigata-u.ac.jp}
\affiliation{Center for Transdisciplinary Research, Niigata University, 8050 Ikarashi-nino-cho, Nishi-ku, Niigata 950-2181, Japan}

\author{Akinori Awazu}
\email[]{awa@hiroshima-u.ac.jp}
\affiliation{Department of Mathematical and Life Sciences, Hiroshima University, 1-3-1, Kagamiyama, Higashi-Hiroshima 739-8526, Japan}

\author{Hiraku Nishimori}
\email[]{nishimor@hiroshima-u.ac.jp}
\affiliation{Department of Mathematical and Life Sciences, Hiroshima University, 1-3-1, Kagamiyama, Higashi-Hiroshima 739-8526, Japan}

\date{\today}

\begin{abstract}
To understand the process of pattern formation in a low-density granular flow, we propose a simple particle model.
This model considers spherical particles moving over an inclined flat surface based on three forces:
gravity as the driving force, repulsive force due to particle collision, and drag force as the particle-- interaction through the ambient fluid.
Numerical simulations of this model are conducted in two different types of two-dimensional planes, i.e., the monolayer was treated.
In the horizontal plane parallel to the slope, particles aggregate at the moving front of the granular flow; and subsequently, flow instability occurs as a wavy pattern.
This flow pattern is caused by the interparticle interaction arising from the drag force.
Additionally, a vortex convection of particles is formed inside the aggregations.
Meanwhile, in the vertical plane on the slope, particle aggregation is also found at the moving front of the granular flow.
The aggregation resembles a head--tail structure, where the frontal angle against the slope approaches 60$^\circ$ from a larger angle as time progresses.
Comparing the numerical result by varying the particle size, the qualitative dynamics of the granular flow are independent of size.
To elucidate this reason, we perform a nondimensionalization for this model.
The result indicates that our model can be simplified to dimensionless equations with one dimensionless parameter that represents the ratio of the gravity term to the excluded volume effect.
\end{abstract}


\maketitle

\section{Introduction}
\label{sec:intro}

Avalanches in nature are generally regarded as a class of massive landslide phenomena involving gravity and density currents.
Their typical granular flows include debris flows, pyroclastic flows, snow avalanches, etc.
These flows slide down a slope as a mixture of solid and fluid, and exhibit various patterns and complicated inner structures.
For instance, powder snow avalanches are divided into two regions for the structure according to field observations~\cite{SOVILLA201597, kohler2016dynamics}: dense flow and powder cloud.
The former is formed in the vicinity of a slope, whereas the latter develops above the dense part.
In general, the dense flow consists of coarse particles or high-density particles that are packed densely in the flow.
Meanwhile, the powder cloud comprises fine particles or low-density particles that are suspended in the air or liquid.
The driving force of both flows is gravity; however, the resisting force is different: inner and basal frictions (dense flow) and drag force (powder cloud).
Hence, these resisting forces lead to a major difference in the structure and dynamics.
A number of laboratory experiments on dense flow have been performed in physics~\cite{Pouliquen:1997aa, forterre2001longitudinal, gray_ancey_2009, gray2011multi} and geophysics~\cite{iverson1997physics, tiefenbacher2004experimental, fischer2018heat}.
Dense granular flows consisting of monodisperse glass beads form the fingering pattern at the front and the streaky structure at the surface~\cite{Pouliquen:1997aa, forterre2001longitudinal}; thus, flow instability occurs.
In these flow patterns, the recirculation of particles at the front is caused by the velocity profile with flow height, and granular convection is confirmed inside the streaky structure.
Moreover, polydisperse particles result in segregation depending on the particle size~\cite{gray_ancey_2009, gray2011multi}.
Because basal friction is a key factor to understand the mechanism of dense flow, it is measured in experiments with snow particles~\cite{tiefenbacher2004experimental}.
Numerical models on dense flow are categorized into two primary types: discrete element method (: DEM)~\cite{PhysRevLett.67.1751, PhysRevE.64.051302} and continuum model~\cite{savage_hutter_1989, forterre2002stability, pitman2003computing, patra2005parallel, el2009two, gray_ancey_2009, gray2011multi}.
Particularly, the continuum approximation of granular matters performs well by utilizing the results of DEM simulations as the constitutive law of the model.
Recently, the granular continuum model is combined with the governing equation for the fluid, and a two-phase flow model has been developed for wet granular flows~\cite{el2009two}.
Meanwhile, experiments on the powder cloud have been conducted in various materials, situations, and scales to adjust several dimensionless numbers~\cite{beghin1981gravitational, beghin1991experimental, nishimura1998ping, mcelwaine2001ping, turnbull2008experiments, nohguchi2009vortex, npg-20-121-2013, jackson2017identification}.
For the particle Reynolds number $Re_\mathrm{p}$, the typical value of the natural powder cloud in snow avalanches is $Re_\mathrm{p} \approx 3000$.
This value is higher than that of the experiments (see Ref.~\cite{turnbull2008experiments}); however, polystyrene--air flows with $Re_\mathrm{p} \approx 150$ can reproduce the cloud pattern except for particles flown by a turbulent eddy~\cite{turnbull2008experiments, nohguchi2009vortex, npg-20-121-2013, jackson2017identification}.
In these experiments, the head--tail structure and the wavy pattern due to flow instability were formed.
Here, the head means a large dense particle cluster at the moving front of the granular flow, whereas the tail means the thin layer behind the head.
The recirculation of particles occurs at the head~\cite{nohguchi2009vortex, jackson2017identification}, and the direction is opposite to the recirculation of dense flow~\cite{Pouliquen:1997aa}.
These dynamics are observed at the actual-scale granular flow consisting of low-density ping-pong balls ($\rho_\mathrm{p} \approx 11~\mathrm{kg~m^{-3}}$)~\cite{nishimura1998ping, mcelwaine2001ping}.
Furthermore, the similar cloud is confirmed at falling particles in a viscous fluid ($Re_\mathrm{p} \ll 1$), although the particles do not move on the slope~\cite{metzger2007falling}.
Many numerical models and theories on powder cloud have been developed as a solid--gas two-phase flow model~\cite{beghin1991experimental, issler1998modelling, naaim1998two, hartel2000analysis1, hartel2000analysis2, ancey2004powder, mcelwaine2005rotational, turnbull2010potential, espath2015high} as the particle is  affected significantly by the fluid.
This type of model is suitable for the large-scale dynamics of an actual mountain topography.
However, any experiments on powder cloud use low-density particles such as polystyrene particles and ping-pong balls.
Hence, a different approach such as the particle-based model is required to understand the mechanism of the powder cloud.
According to Ref.~\cite{nishimura1998ping}, they developed a numerical model that coupled between the DEM for the particles and the Reynolds-averaged Navier--Stokes equations for the fluid.
However, this model could not reproduce the flow pattern of the experiments; thus, they suggested considering the strong interaction between each particle trajectory and the fluid.
In this study, we propose a simple particle model for the granular flow consisting of low-density particles to qualitatively understand the pattern formation and each particle trajectory that interacted with the ambient fluid.
In Section~\ref{sec:method}, we describe the numerical modeling and subsequently explain the setup of the numerical simulations with the model.
In Section~\ref{sec:results}, simulation results regarding flow pattern and speed are presented.
In Section~\ref{sec:discussions}, the shapes of the flow patterns are discussed.


\section{Method}
\label{sec:method}

\subsection{Concept of model}
\label{subsec:concept}

\begin{figure}[t]
\includegraphics[width=1\linewidth]{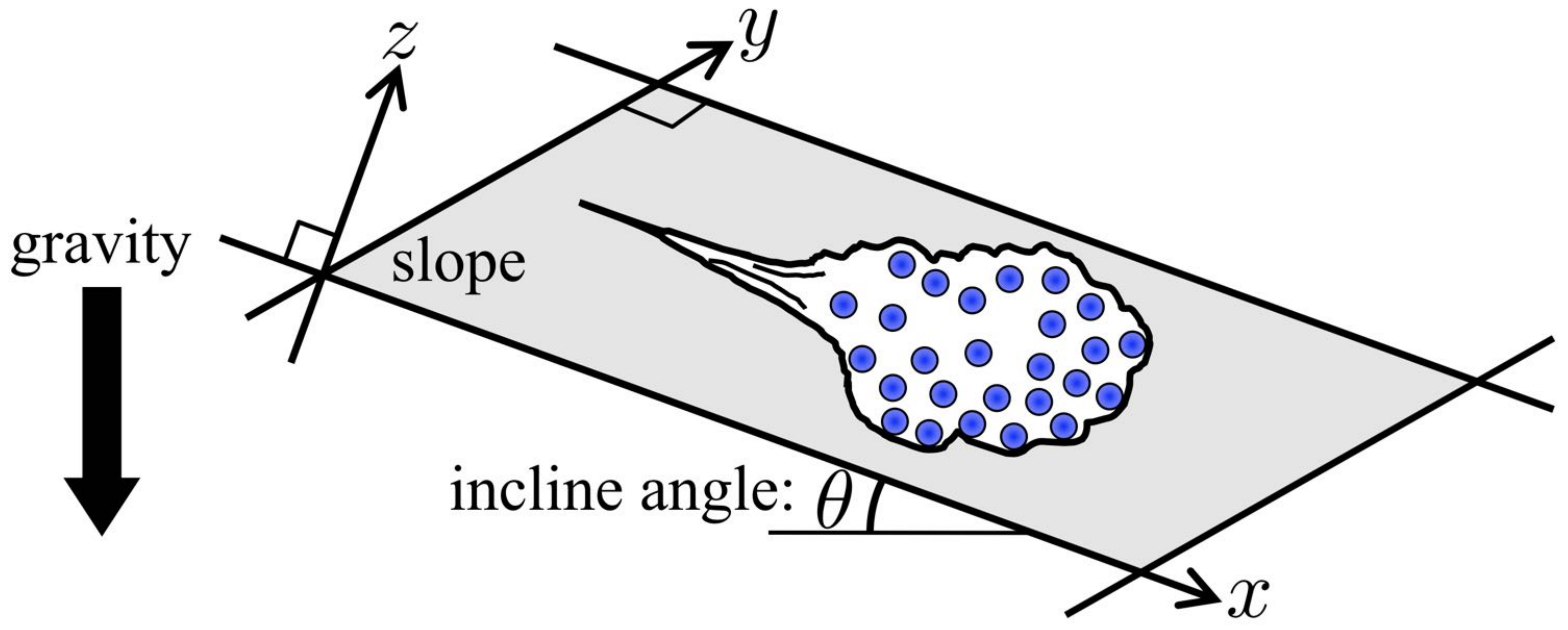}%
\caption{
Schematic image of simple particle model.
The $x$-$y$ plane is parallel to the slope with the incline angle of $\theta$ ($x$: inclination direction, $y$: lateral direction).
The $z$-axis is perpendicular to the slope.
\label{fig:schematic}}
\end{figure}

We propose a simple particle model for low-density granular flow based on the equation of motion for particles.
Realistically, the granular flow of natural phenomena is affected by many physical factors: gravity, basal and inner frictions, consolidation, laminar and turbulent flows, air drag, etc.
Thus, the dynamics of granular flow becomes complex and the understanding of the mechanism becomes difficult.
Our model aims to significantly simplify the complicated behavior of low-density granular flows interacting with the ambient fluid.
Our model is not realistic yet; however, we aim to reproduce the flow pattern found in previous experiments with ping-pong balls and polystyrene particles~\cite{mcelwaine2001ping,nohguchi2009vortex}.
In our model, a low-density granular flow slides on a slope at a steeper angle than the angle of repose (Fig.~\ref{fig:schematic}).
This model is based on three basic assumptions for the simplification of modeling:
(i) The granular flow consists of hard-sphere particles with a low true density.
(ii) Only the translational motion of particles is considered, i.e., the rotational motion is ignored.
(iii) Three types of forces act on the particles, i.e.,
gravity as the driving force for granular flow, repulsive force between particles as the excluded volume effect, and drag force due to the ambient fluid.
Figure~\ref{fig:schematic} shows the schematic image of this model, where $\theta$ is the incline angle, $x$-axis is the inclination direction, $y$-axis is the lateral direction, and $z$-axis is perpendicular to the slope.

\subsection{Numerical modeling}
\label{subsec:numerical}

\begin{figure}[t]
\includegraphics[width=1\linewidth]{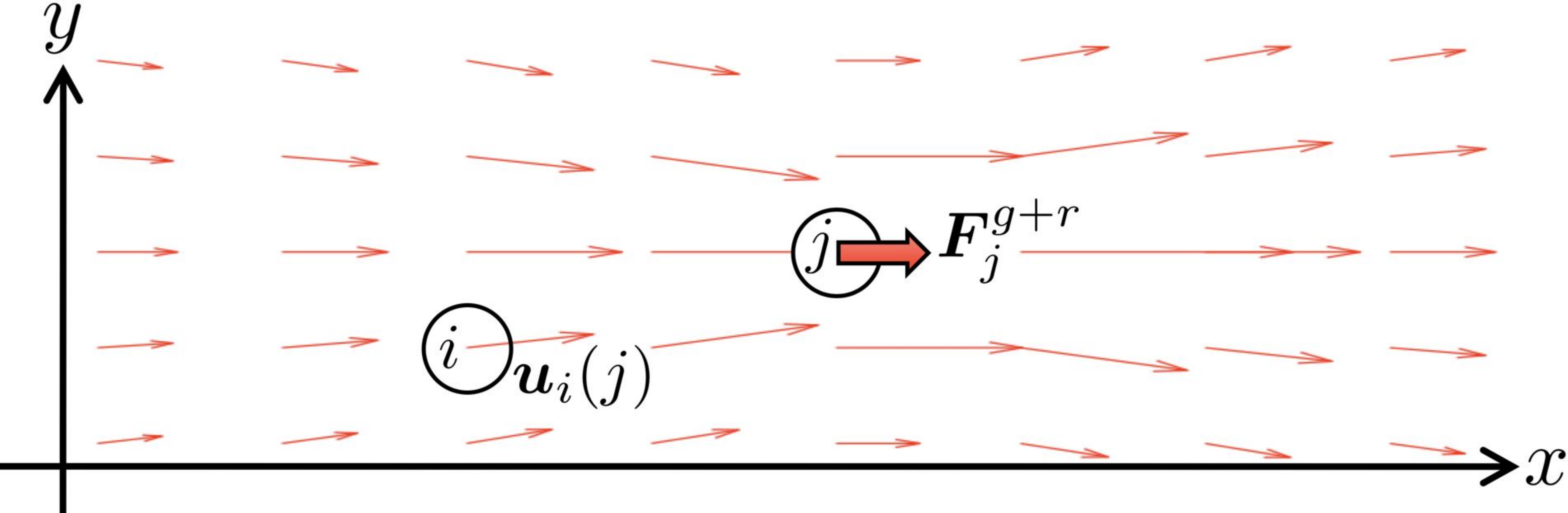}%
\caption{
Induced velocity field $\bm{u}(j)$ generated by gravity and repulsive force of the $j$th particle, $\bm{F}_j^{g + r}$, using Eq.~(\ref{subeq:drag2}).
According to the force direction, the ambient fluid is pushed and pulled as shown by arrows.
The induced velocity at the $i$th particle is expressed as $\bm{u}_i(j)$.
\label{fig:induced}}
\end{figure}

To derive the governing equation for the particles, we formulate the force acting on the $i$th particle $\bm{F}_i = \bm{F}_i^g + \bm{F}_i^r + \bm{F}_i^d$,
where $\bm{F}_i^g$ is gravity, $\bm{F}_i^r$ is the repulsive force between particles, and $\bm{F}_i^d$ is the drag force due to the ambient fluid.
Hereinafter, the radius, coordinate, and velocity of the $i$th particle are denoted by $a_i$, $\bm{r}_i$, and $\bm{v}_i$, respectively, and the true density of particles is denoted by $\rho_\mathrm{p}$.
Gravity $\bm{F}^g$:
The gravity in the vertical direction of the $i$th particle $F_i^g$ is calculated by considering the density difference between the particle and fluid;
subsequently, it is divided into $x$ and $z$ components depending on the incline angle $\theta$, as follows:
\begin{subequations}
\label{eq:gravity}
\begin{eqnarray}
\bm{F}_i^g
&=&
F_i^g
\left(
\begin{array}{c}
- \sin \theta \\
0 \\
\cos \theta
\end{array}
\right)
=
F_i^g e_\theta,
\label{subeq:gravity1}
\\
F_i^g
&=&
- V_i g (\rho_\mathrm{p} - \rho_\mathrm{f}),
\label{subeq:gravity2}
\\
V_i
&=&
\frac{4}{3} \pi a_i^3,
\label{subeq:gravity3}
\end{eqnarray}
\end{subequations}
where $e_\theta$ is the vector, $V_i$ is the volume of the $i$th particle, $g$ is the gravitational acceleration, and $\rho_\mathrm{f}$ is the density of the fluid.
Repulsive force $\bm{F}^r$:
Because the rotational motion of the particles is ignored, only the normal force of collision between the $i$th and $j$th particles is calculated as the repulsive force of the $i$th particle $\bm{F}_i^r$ (and $\bm{F}_j^r$), which is represented by the elastic spring force $F_i^r$ as follows:
\begin{subequations}
\label{eq:repulsive}
\begin{eqnarray}
\bm{F}_i^r
&=&
F_i^r
\frac{\bm{r}_j - \bm{r}_i}{r_{ij}}
=
F_i^r \bm{n}_{ij}
,
\label{subeq:repulsive1}
\\
F_i^r
&=&
- k_n \delta_{ij},
\label{subeq:repulsive2}
\\
\delta_{ij}
&=&
\left\{
\begin{array}{cl}
a_i + a_j - r_{ij} & (r_{ij} \le a_i + a_j: \text{contact}) \\
0 & (r_{ij} > a_i + a_j: \text{non-contact}).
\end{array}
\right.
\label{subeq:repulsive3}
\end{eqnarray}
\end{subequations}
Here, $r_{ij} = |\bm{r}_j - \bm{r}_i|$ is the interparticle distance and $\bm{n}_{ij}$ is the unit vector joining the centers of two particles.
Further, $k_n$ is the linear spring constant and $\delta_{ij}$ is the overlap between two particles.
We calculate the collision between the particles and the slope using Eq.~(\ref{eq:repulsive}).
Drag force $\bm{F}^d$:
To estimate the exact drag force acting on the particles such as the air drag, the computational fluid dynamics of the interacting particles are required.
In this modeling, we prioritize the qualitative reproduction of drag force over the realistic one.
Concretely, the ambient fluid is assumed to be Stokes flow (i.e., low Reynolds number $Re \ll 1$), although the particle inertia is not negligible and the turbulent eddy occurs in the actual granular flow.
The Stokes flow containing two spherical particles can be solved theoretically when a sufficient distance exists between two particles.
The exact solution is known as the Rotne--Prager tensor $\bm{\mathrm{J}}(\bm{r})$~\cite{rotne1969variational}.
Using this tensor, we express the interparticle interaction through the ambient fluid, that is, the drag force of the $i$th particle $\bm{F}_i^{d}$ is generated by the force acting on the $j$th particle $\bm{F}_j^{g + r} = \bm{F}_j^{g} + \bm{F}_j^{r}$, as follows:
\begin{subequations}
\label{eq:drag}
\begin{eqnarray}
\bm{F}_i^d
&=&
6 \pi \mu a_i \bm{u}_i(j),
\label{subeq:drag1}
\\
\bm{u}_i(j)
&=&
\frac{1}{8 \pi \mu} \bm{\mathrm{J}}(\bm{r} = \bm{r}_i - \bm{r}_j) \cdot \bm{F}_j^{g+r}
,
\label{subeq:drag2}
\\
\bm{\mathrm{J}}(\bm{r})
&=&
\frac{1}{r_{ji}}
\left[
\bm{I} + \frac{\bm{r} \bm{r}}{r_{ji}^2}
+ \frac{2}{3} \left( \frac{a_j}{r_{ji}} \right)^2
\left( \bm{I} - 3 \frac{\bm{r} \bm{r}}{r_{ji}^2} \right)
\right].
\label{subeq:drag3}
\end{eqnarray}
\end{subequations}
In these equations, $\mu$ is the viscosity coefficient of the fluid, $\bm{u}_i (j)$ is the induced velocity by the $j$th particle at the coordinate of the $i$th particle (Fig.~\ref{fig:induced}), and $\bm{I}$ is the unit tensor.
Particularly, $\bm{r} \bm{r}$ in Eq.~(\ref{subeq:drag3}) denotes the tensor corresponding to $\bm{I}$: for instance, $\bm{r} \bm{r}_{xy} = (x_i - x_j) (y_i - y_j)$.
According to the assumption for the drag force (i.e., Stokes flow),
we derive the governing equation for the $i$th particle from Stokes' drag, as follows:
\begin{subequations}
\label{eq:governing}
\begin{eqnarray}
\bm{v}_i
&=&
\frac{\bm{F}_i}{6 \pi \mu a_i}
=
\frac{\bm{F}_i^{g + r}}{6 \pi \mu a_i}
+
\sum_{j \ne i}^{N} \bm{u}_i(j),
\label{subeq:governing1}
\\
\frac{d\bm{r}_i}{dt}
&=&
\bm{v}_i,
\label{subeq:governing2}
\end{eqnarray}
\end{subequations}
where $N$ is the number of particles.
In Eq.~(\ref{subeq:governing1}), the second term indicates the long-range interaction between particles through the ambient fluid, which resembles the asymmetric potential (Fig.~\ref{fig:induced}).

\subsection{Simulation setup}
\label{subsec:simulation}

\begin{figure*}[t]
\includegraphics[width=1.0\linewidth]{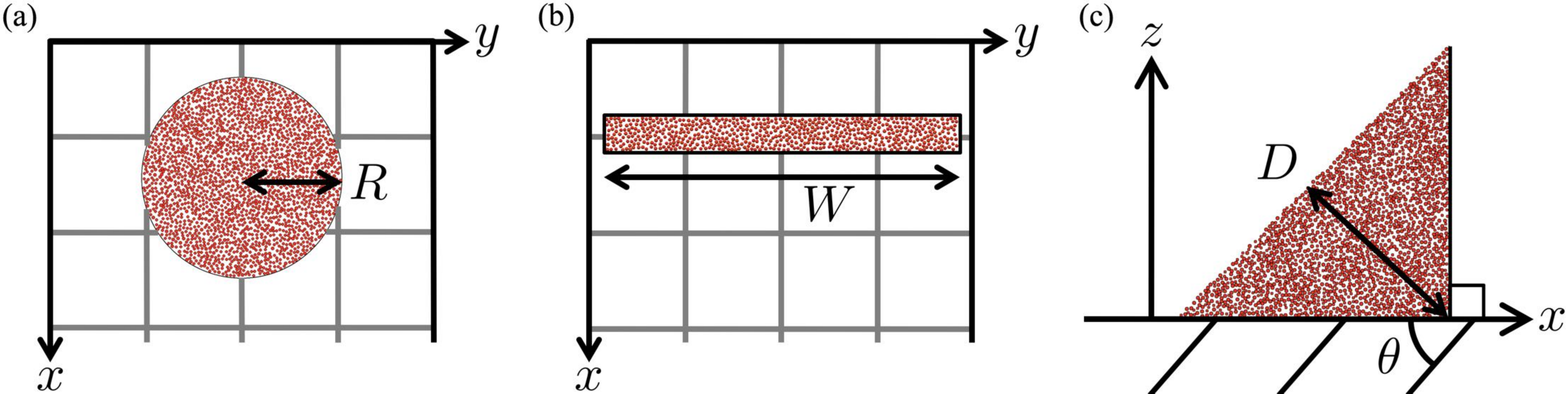}%
\caption{
Three different types of initial conditions for particles in two-dimensional planes are as follows:
(a) circle in $x$--$y$ plane, (b) rectangle in $x$--$y$ plane, and (c) triangle in $x$--$z$ plane.
The $x$--$y$ plane is parallel to the slope, whereas the $x$--$z$ plane is the vertical cross-section on the slope, where $\theta = 45^\circ$ is the incline angle.
In each initial condition, the circle radius $R$,  rectangle width $W$, and triangle depth $D$ are as constant values depending on the particle size, such that the particle volume fraction $\Phi_\mathrm{p}$ of 2000 particles is approximately 0.5.
\label{fig:three}}
\end{figure*}
The numerical modeling for the low-density granular flow was performed in a three-dimensional space, as shown in Fig.~\ref{fig:schematic}; however, we used the time complexity of $O(N^2)$ in our model to calculate the drag force $\bm{F}_i^d$ in Eq.~(\ref{eq:drag}).
The repulsive force $\bm{F}_i^r$ in Eq.~(\ref{eq:repulsive}) additionally requires a small timestep $dt$ used in the numerical simulation.
Hence, this model is unsuitable for the calculation with a large number of particles.
In this study, the number of particles is set as $N=2000$; subsequently, numerical simulations were conducted in two different types of two-dimensional spaces (i.e., monolayer);
The $x$--$y$ plane is parallel to the slope, while the $x$--$z$ plane is perpendicular to the slope.
Figure~\ref{fig:three} shows three different types of initial conditions for the particles in the $x$--$y$ and $x$--$z$ planes:
(a) circular shape, (b) rectangular shape, and (c) triangular shape.
Here, the incline angle is fixed as $\theta = 45^\circ$.
These shapes are determined by reference to previous experiments with polystyrene particles~\cite{nohguchi2009vortex}, although the size is small in comparison to Ref.~\cite{nohguchi2009vortex} because of the small $N$.
The circle radius $R$, rectangle width $W$, and triangle depth $D$ are set as constant values depending on the particle size used in the simulations, as will be shown later.
Particularly, in the case of Fig.~\ref{fig:three}(b), the rectangle length in the $x$ direction is fixed as 10 particles.
In each shape, 2000 particles are packed randomly without an overlap.
Regarding the specific materials of the particle and fluid, we assume that the granular flow consists of low-density particles such as polystyrene in air, to compare qualitatively with the experiments~\cite{nohguchi2009vortex}.
The particle density and linear spring constant for the repulsive force are given as $\rho_\mathrm{p} = 20~\mathrm{kg~m^{-3}}$ and $k_n = 10~\mathrm{N~m^{-1}}$, respectively, whereas the density and viscosity coefficient of the fluid are given as $\rho_\mathrm{f} = 1.2~\mathrm{kg~m^{-3}}$ and $\mu = 1.82 \times 10^{-5}~\mathrm{Pa~s}$, respectively.
It is noteworthy that the value of $k_n$ is confirmed, and that the excluded volume effect performs well.
As the control parameter, we varied only the particle radius: $a_\mathrm{p} = \mathrm{1,~2.5,~and~5~mm}$ (three types).
To avoid the crystallization of particles, particles with a polydispersity of $\pm 5~\%$ were used in simulations.
The sizes of the initial shapes $R,~W,~\text{and}~D$, as shown in Fig.~\ref{fig:three} were determined depending on $a_\mathrm{p} = \mathrm{1,~2.5,~and~5~mm}$:
$R = \mathrm{0.065,~0.17,~and~0.34~m}$, $W = \mathrm{0.35,~0.85,~and~1.7~m}$, and $D = \mathrm{0.115,~0.3,~and~0.6~m}$.
These sizes normalized by $a_\mathrm{p}$ are $\tilde{R} = \text{65-68}$, $\tilde{W} = \text{340-350}$, and $\tilde{D} = \text{115-120}$, respectively.
Further, the particle volume fraction is approximately 0.5.
We performed 20 simulations at each initial condition and each particle radius.


\section{Results}
\label{sec:results}

\subsection{Circular shape in $x$--$y$ plane}
\label{subsec:circular}

\begin{figure}[t]
\includegraphics[width=1\linewidth]{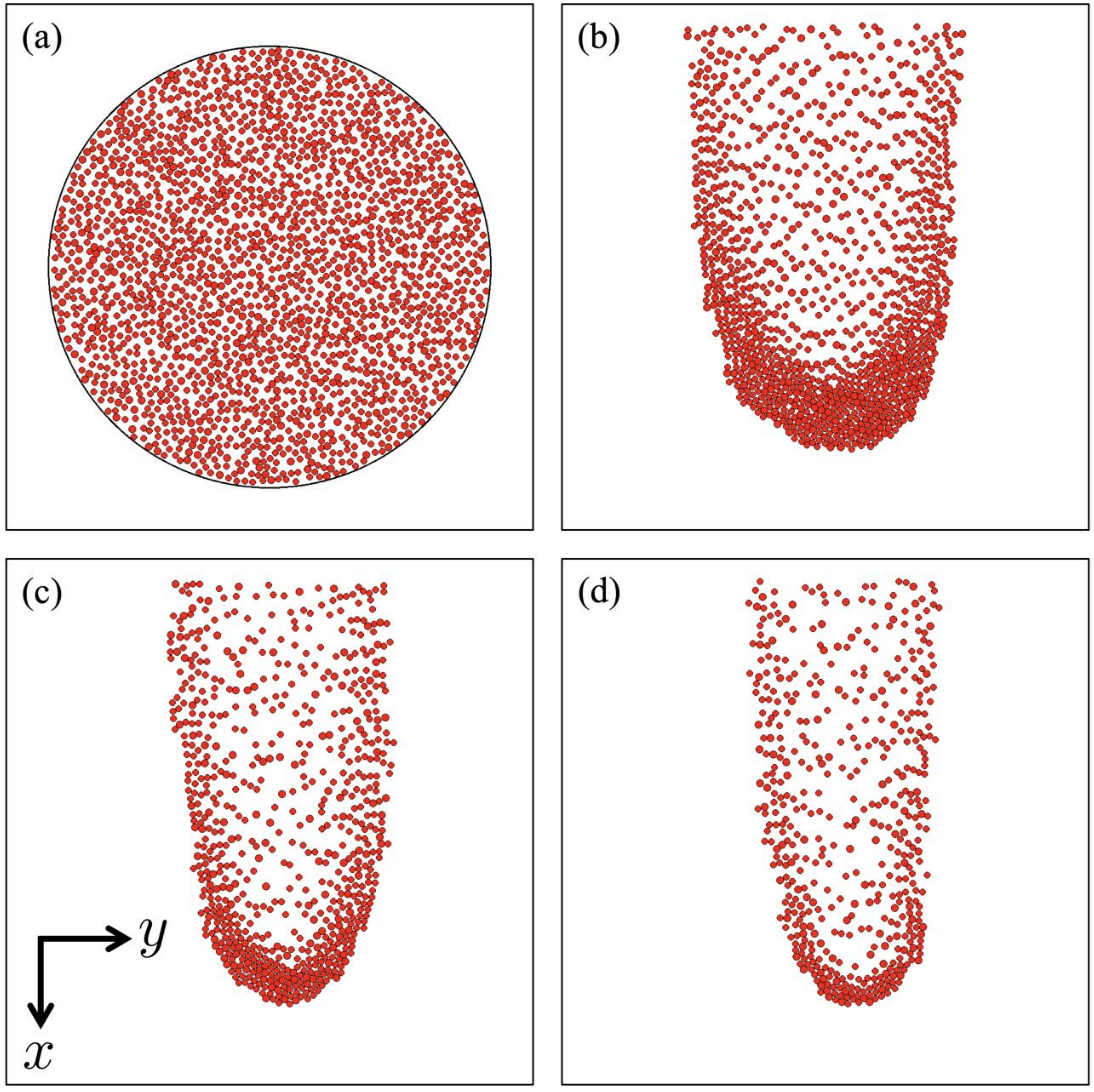}%
\caption{
Time evolution of single head formation with $a_\mathrm{p}$ = 5 mm in circular setup:
(a) $t = 0~\mathrm{s}$,
(b) $t = 10^{-3}~\mathrm{s}$,
(c) $t = 2.45 \times 10^{-3}~\mathrm{s}$, and
(d) $t = 5 \times 10^{-3}~\mathrm{s}$.
\label{fig:time}}
\end{figure}
The typical time evolutions of the granular flow in the circular shape are shown in Fig.~\ref{fig:time}.
Here, the particle radius is $a_\mathrm{p} =$ 5 mm.
In the early stage of the simulations, the particle located at the central part of the circle migrates rapidly to the moving front of the granular flow as time progresses.
Consequently, the initial uniform distribution of particles is shifted to the local distribution at the moving front of the granular flow (Figs.~\ref{fig:time}(a) and (b)).
This pattern of particle aggregation appears to be a single head resembling a crescentic form, similar to the pattern of a low-density granular flow~\cite{mcelwaine2001ping, nohguchi2009vortex}.
After the head formation, the particle is moved away gradually from the aggregation part; subsequently, the head shrinks with time (Figs.~\ref{fig:time}(c) and (d)).
This head-formation process is also confirmed in the other particle radiuses, although the time required for the head formation is different.
To verify the key factor for the head formation, we performed the simulation without the drag force.
In this case, the initial pattern was maintained because the basal friction and particle inertia were ignored in our model.
Therefore, the drag force generated the long-range interaction between particles through the ambient fluid as the induced velocity $\bm{u}_i(j)$ in Eq.~(\ref{eq:drag}), which is essential to form the head.
The reason for the head formation owing to the induced velocity is clear.
In the initial stage of the simulations (Fig.~\ref{fig:time}(a)), the induced velocity is high at the central part because it is inversely proportional to the interparticle distance.
Hence, particle aggregation occurred at the moving front of the granular flow and the aggregation part migrates at a faster speed.
\begin{figure}[t]
\includegraphics[width=0.8\linewidth]{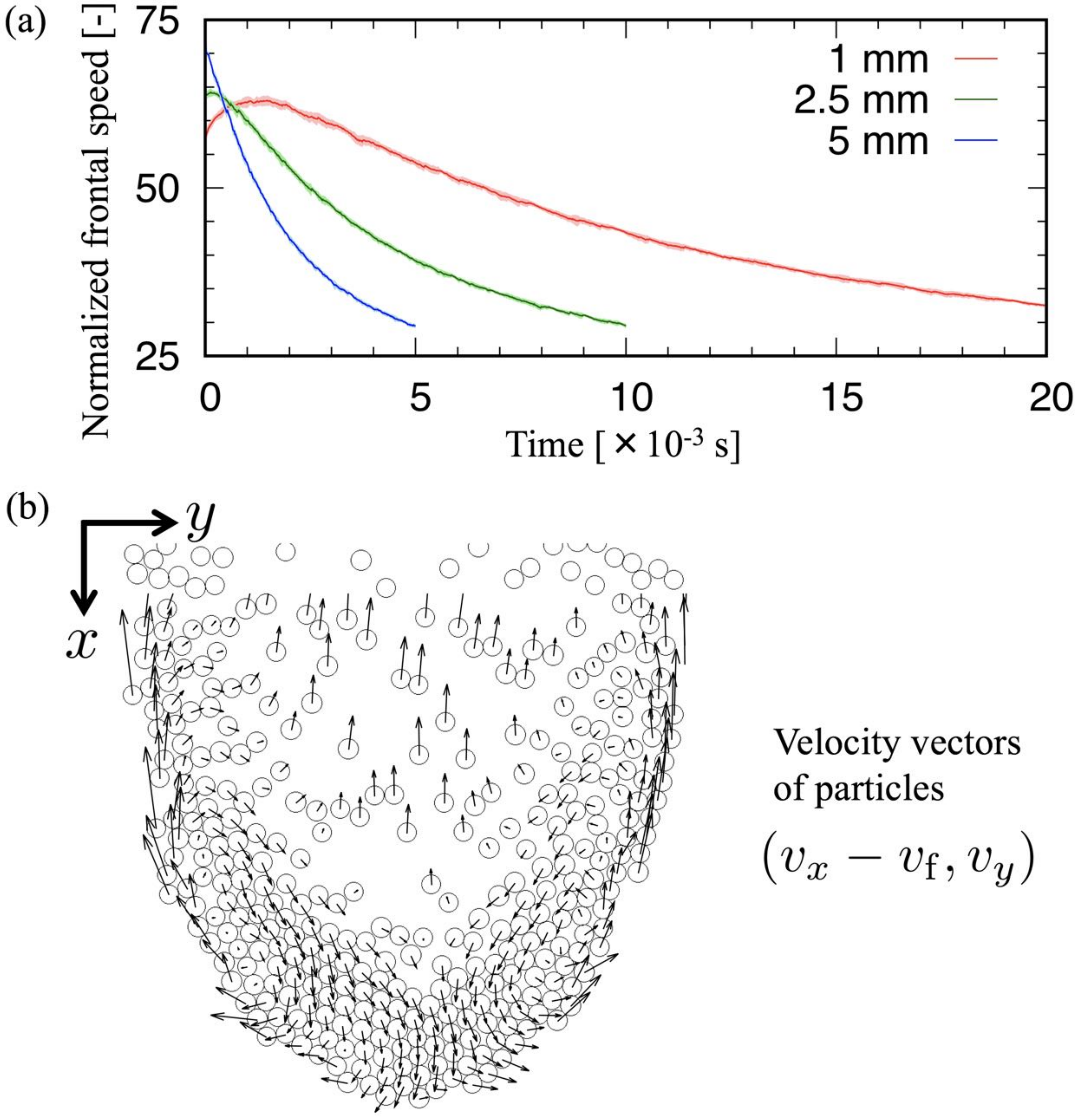}%
\caption{
Temporal variations of particle velocity in circular setup:
(a) normalized frontal speed of single head $\tilde{v}_\mathrm{f}$ at different particle radiuses $a_\mathrm{p}$ and (b) velocity vector of particles at the head, as shown in Fig. \ref{fig:time}(c).
(a)
The mean and standard deviation of 20 simulations are denoted by a solid curve and thin band, respectively.
\label{fig:temporal}}
\end{figure}
For the migration speed of the granular flow, we measure the front speed in the $x$ direction $v_\mathrm{f}$ according to the displacement of the forefront of the granular flow.
In our model, the speed of a single particle in the $x$ direction $v_\mathrm{s}$ depends significantly on the particle radius $a_\mathrm{p}$, as follows:
\begin{eqnarray}
v_\mathrm{s}
=
\frac{2}{9} \frac{g (\rho_\mathrm{p} - \rho_\mathrm{f}) \sin \theta}{\mu} a_\mathrm{p}^2.
\label{eq:single}
\end{eqnarray}
Consequently, the frontal speed is normalized by the single-particle speed as $\tilde{v}_\mathrm{f} = v_\mathrm{f} / v_\mathrm{s}$.
Because $v_\mathrm{s}$ increases with $a_\mathrm{p}$, the smaller time step $dt$ is required for the calculation of repulsive force in the larger $a_\mathrm{p}$.
Numerical simulations continue until the granular flow reaches the state shown in Fig.~\ref{fig:time}(d):
$t_\mathrm{end} = \text{20,~10,~and~5~ms}~(a_\mathrm{p} = \text{1,~2.5,~and~5~mm})$.
Figure~\ref{fig:temporal}(a) shows the temporal variations of normalized frontal speeds $\tilde{v}_\mathrm{f} (t)$ measured at different particle radiuses $a_\mathrm{p}$.
The solid curve and thin band denote the mean and standard deviation of 20 simulations, respectively.
The standard deviation is small against the mean, even though the size distribution and initial position of the particles are changed.
In all cases, $\tilde{v}_\mathrm{f}$ increases significantly in the early stage and subsequently decreases gradually with time because of the head shrinkage.
Additionally, $\tilde{v}_\mathrm{f}$ ranges from 30 to 70 independent of $a_\mathrm{p}$, although the change rate depends on $a_\mathrm{p}$.
Therefore, the head size (i.e., number of particles) results in the change in $\tilde{v}_\mathrm{f}$ (Figs.~\ref{fig:time} and \ref{fig:temporal}(a)).
The head size is discussed in detail in  Section~\ref{subsec:characteristics}.
The movements of particles at the head can be visualized by the velocity vectors of the particles (Fig.~\ref{fig:temporal}(b)).
This particle location corresponds to that in Fig.~\ref{fig:time}(c), that is, $t =$ 2.45 ms and $a_\mathrm{p} =$ 5 mm.
To study the particle velocity relative to the frontal speed of the granular flow $v_\mathrm{f}$, the $x$ component of the vectors is denoted by $v_x - v_\mathrm{f}$, where $v_x$ is the $x$ component of the particle velocity.
Consequently, a granular vortex convection was found at the head and particles away from the head.
Particularly, the particle at the rear of the head moves inward to the forefront of the head, whereas the particle at the front of the head is pushed outward.
Some particles return to the head again; however, most of the particles are left behind.
Thus, the head shrinks with time.

\subsection{Rectangular shape in $x$--$y$ plane}
\label{subsec:rectangular}

\begin{figure}[t]
\includegraphics[width=1\linewidth]{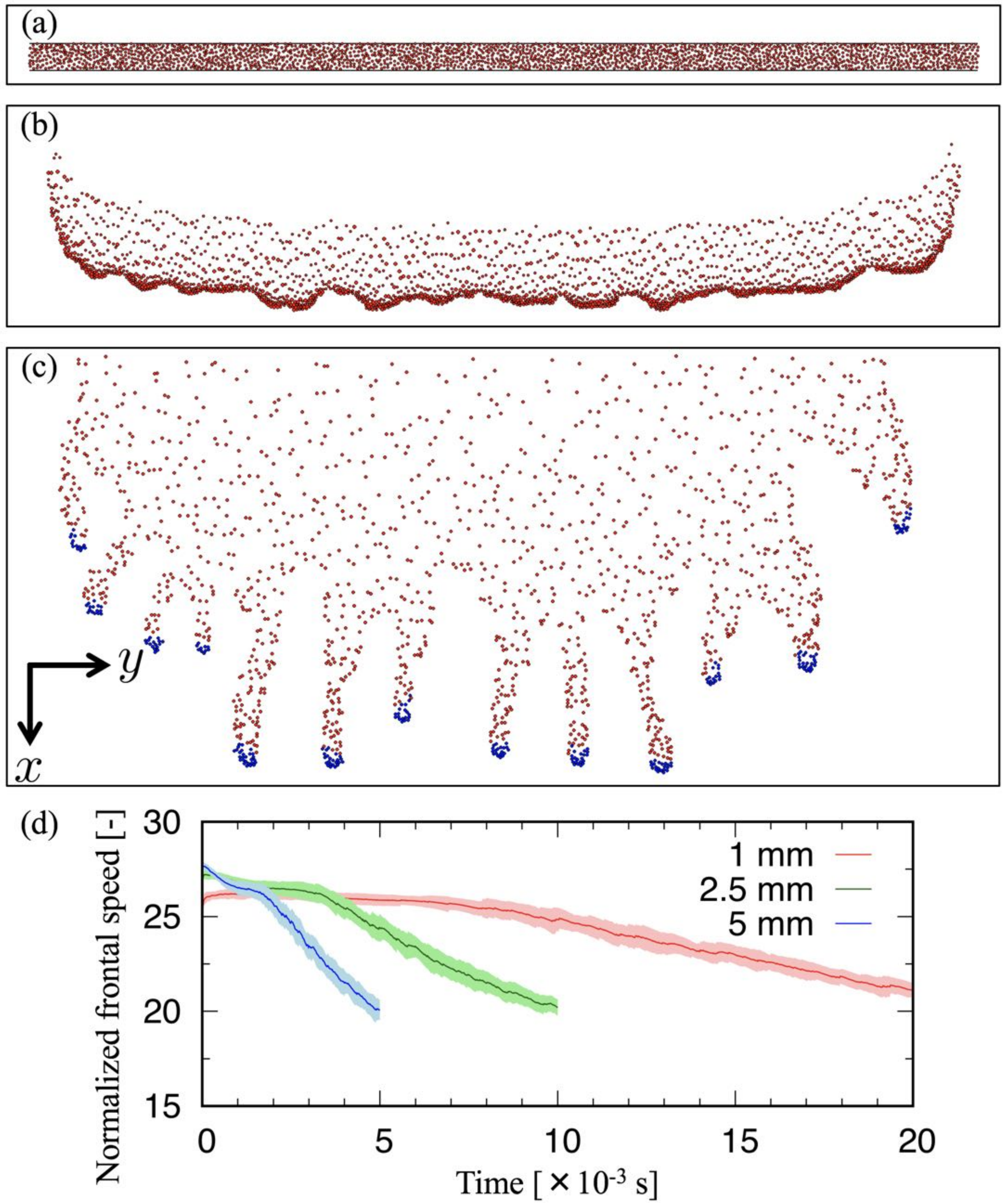}%
\caption{
Flow instability from straight shape to wavy pattern with $a_\mathrm{p}$ = 5 mm in rectangular setup:
(a) $t = 0~\mathrm{s}$,
(b) $t = 10^{-3}~\mathrm{s}$, and
(c) $t = 5 \times 10^{-3}~\mathrm{s}$.
(c)
The blue particle is the aggregation part detected as a head, according to Eq. (\ref{eq:index}).
(d)
Three normalized frontal speeds $v_\mathrm{f}(t)$ are shown at $a_\mathrm{p}$ = 1, 2.5, and 5 mm.
The mean and standard deviation of 20 simulations are denoted by solid curve and thin band, respectively.
\label{fig:flow}}
\end{figure}
The typical time evolutions of granular flow in the rectangular shape are shown in Figs.~\ref{fig:flow}(a), (b), and (c).
Here, the particle radius is $a_\mathrm{p} =$ 5 mm.
The simulation time at each $a_\mathrm{p}$ is set the same as that in the circular shape:
$t_\mathrm{end} = \text{20,~10,~and~5~ms}~(a_\mathrm{p} = \text{1,~2.5,~and~5~mm})$.
In the early stage of the simulations, the initial straight shape at the moving front of the granular flow is maintained, although the particle  moves forward slowly to the front by the long-range interaction between the particles due to the drag force.
Subsequently, the straight shape with particle aggregation deforms into a wavy pattern (Fig.~\ref{fig:flow}(b)), that is, flow instability occurs.
This destabilization of the moving front is caused by the inhomogeneous distribution of the size and coordinate in the initial condition of the particles (Fig.~\ref{fig:flow}(a)).
Because the aggregation part is faster than the other parts, the wavy pattern elongates in a downward direction.
Eventually, the multihead structure is formed from the straight shape (Fig.~\ref{fig:flow}(c)).
Figure~\ref{fig:flow}(d) shows the front speed in the $x$ direction normalized by the single-particle speed, $\tilde{v}_\mathrm{f}$, at each particle radius $a_\mathrm{p}$.
The mean and standard deviation of 20 simulations are denoted by the solid curve and thin band, respectively.
In all cases, $\tilde{v}_\mathrm{f}$ decreases gradually in the early stage; subsequently, the decreasing rate of $\tilde{v}_\mathrm{f}$ tends to be relatively high.
Additionally, $\tilde{v}_\mathrm{f}$ ranges from 20 to 28 independent of $a_\mathrm{p}$, and the multihead structure is confirmed at the end of each simulation (Fig.~\ref{fig:flow}(c)).
The formed pattern (number of heads, head size) is different depending on the initial condition; however, the wavelength reproducibility is good.
This point is discussed in detail in  Section~\ref{subsec:characteristics}.
In Fig.~\ref{fig:flow}(d), the rate of decrease in $\tilde{v}_\mathrm{f}$ is linked with the flow pattern, as shown in Figs.~\ref{fig:flow}(a), (b), and (c).
For example, during the deformation from the straight shape (Fig.~\ref{fig:flow}(a)) to the wavy pattern (Fig.~\ref{fig:flow}(b)), the number of particles constituting the aggregation part is nearly unchanged.
Thus, $\tilde{v}_\mathrm{f}$ is maintained in the early stage of the simulations.
Meanwhile, the number of particles decreases locally with the elongation of heads from the wavy pattern.
Consequently, $\tilde{v}_\mathrm{f}$ decreases at a relatively high rate since the middle stage of the simulations.

\subsection{Triangular shape in $x$--$z$ plane}
\label{subsec:triangular}

\begin{figure}[t]
\includegraphics[width=1\linewidth]{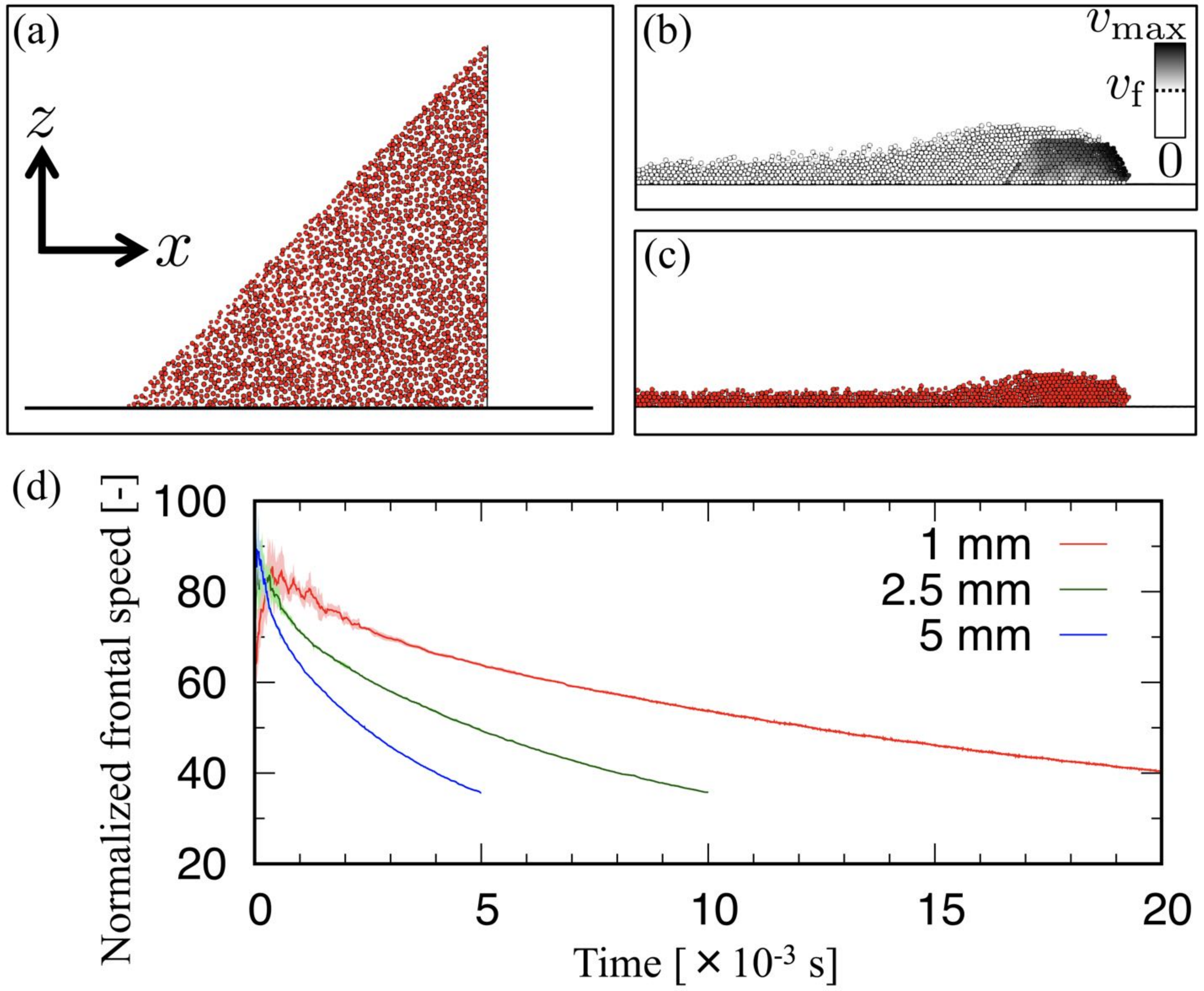}%
\caption{
Formation of head--tail structure with $a_\mathrm{p}$ = 5 mm in triangular setup:
(a) $t = 0~\mathrm{s}$,
(b) $t = 2 \times 10^{-3}~\mathrm{s}$, and
(c) $t = 5 \times 10^{-3}~\mathrm{s}$.
(b)
The gray scale denotes the $x$ component of the particle velocity.
The white particles are slower compared to the frontal speed $v_\mathrm{f}$, whereas the gray particles are faster compared to $v_\mathrm{f}$.
Here, $v_\mathrm{f}$ is calculated from the particle on the slope (i.e., first layer).
(d)
Three normalized frontal speeds $v_\mathrm{f}(t)$ are shown at $a_\mathrm{p}$ = 1, 2.5, and 5 mm.
The mean and standard deviation of 20 simulations are denoted by a solid curve and thin band, respectively.
\label{fig:formation}}
\end{figure}
The typical time evolutions of the granular flow in the triangular shape are shown in Figs.~\ref{fig:formation}(a), (b), and (c).
Here, the incline angle and particle radius are $\theta = 45^{\circ}$ and $a_\mathrm{p} =$ 5 mm, respectively.
The simulation time at each $a_\mathrm{p}$ is set as $t_\mathrm{end} = \text{20,~10,~and~5~ms}~(a_\mathrm{p} = \text{1,~2.5,~and~5~mm})$.
Because the particles are up in the air initially (Fig.~\ref{fig:formation}(a)), they immediately accumulate near the slope according to the gravity.
Subsequently, a large-scale cluster consisting of particles emerges from the accumulation, and moves down a slope at a high speed (Fig.~\ref{fig:formation}(b)).
This pattern appears to be the head--tail structure, where the thickness in the $z$ direction of the front part is more than twice that of the rear part.
This profile is similar to the flow height of the experiments with ping-pong balls~\cite{nishimura1998ping}.
After the head--tail structure is formed, the head shrinks with time (Fig.~\ref{fig:formation}(c)).
In the $x$--$z$ plane, the front speed in the $x$ direction $v_\mathrm{f}$ is measured according to the displacement of the forefront particle on the slope (i.e., first layer).
Figure~\ref{fig:formation}(d) shows the normalized front speed $\tilde{v}_\mathrm{f} = v_\mathrm{f}/v_\mathrm{s}$ at different particle radiuses $a_\mathrm{p}$.
The solid curve and thin band denote the mean and standard deviation of 20 simulations, respectively.
In each case, $\tilde{v}_\mathrm{f}$ increases significantly with the cluster formation in the early stage; subsequently, $\tilde{v}_\mathrm{f}$ decreases with time because of the head shrinkage.
The standard deviation is negligibly small after the head--tail structure is formed; this implies that the identical structure is formed independently of the initial condition.
To verify the particle movement in the head--tail structure, we focus on the $x$ component of the particle velocity.
In Fig.~\ref{fig:formation}(b), the white particle is slower compared to the front speed $v_\mathrm{f}$, whereas the gray particle is faster compared to $v_\mathrm{f}$.
The particle speed at the front part of the granular flow exceeds $v_\mathrm{f}$ except the upper part of the head, whereas that at the rear part (i.e., tail) is smaller than $v_\mathrm{f}$.
Therefore, the particle moves from the upper part of the head to the tail.
These particle movements lead to the head shrinkage.


\section{Discussions}
\label{sec:discussions}

\subsection{Definition of head in $x$--$y$ plane}
\label{subsec:definition}

\begin{figure}[t]
\includegraphics[width=1\linewidth]{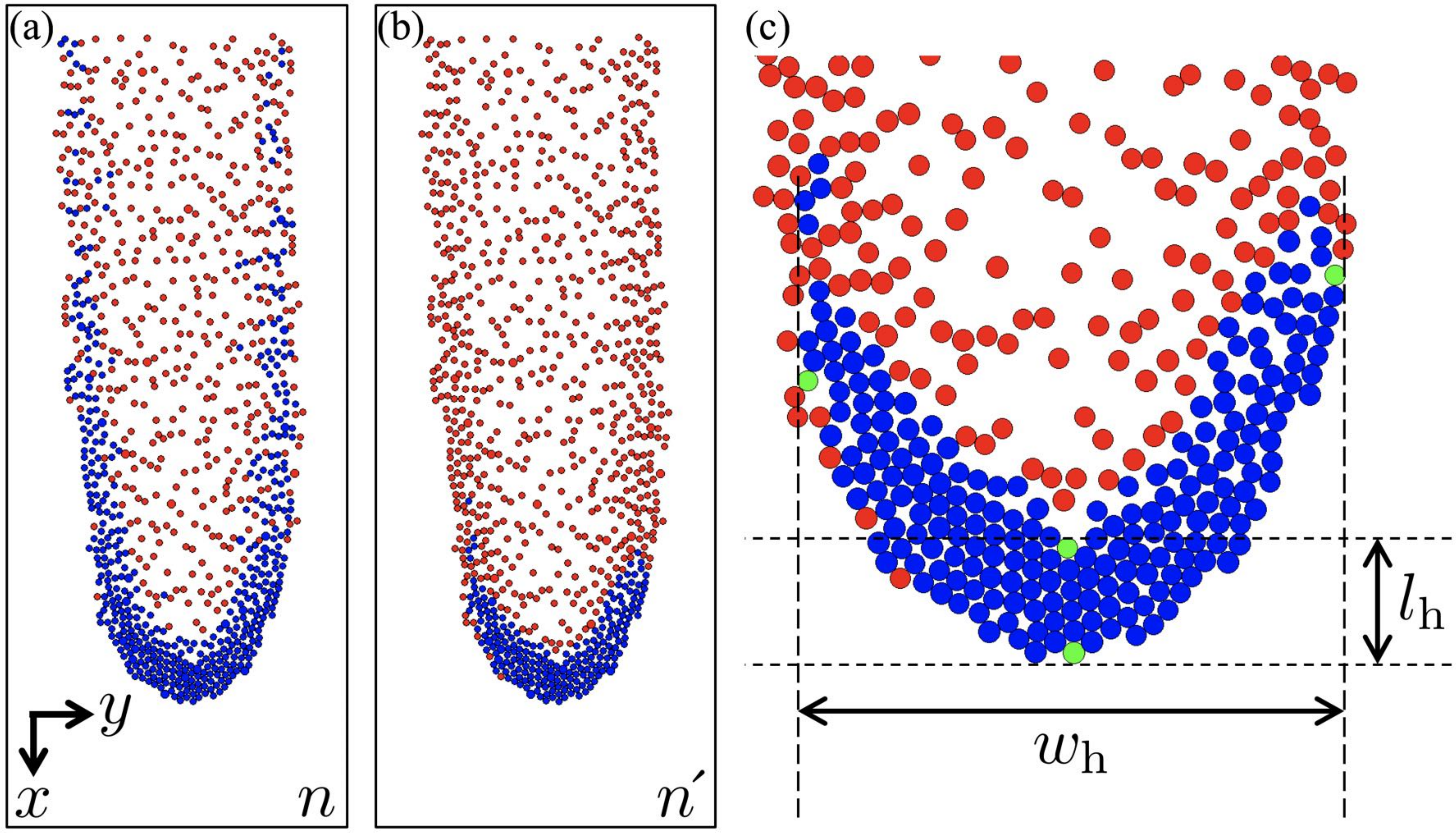}%
\caption{
Examples of head in $x$--$y$ plane defined according to Eq. (\ref{eq:index}):
(a) $n$, (b) and (c) $n\acute{}$.
The blue and green particles constitute a head at the front of the granular flow in which four green particles are selected as targets to measure the width $w_\mathrm{h}$ and layer thickness $l_\mathrm{h}$ of the head.
\label{fig:examples}}
\end{figure}
As mentioned in Sections~\ref{subsec:circular} and \ref{subsec:rectangular}, the front speed of the granular flow relates to the head size.
The multihead structure is formed from the straight shape in the $x$--$y$ plane, which is parallel to the slope.
Hereinafter, we define the head based on the particle number density to discuss the temporal variation of the head size and the size distribution of the head.
Our idea is to detect the particle of relatively high particle number density from the flow pattern, as shown in Fig.~\ref{fig:examples}.
As a simple index, we count two different types of numbers around the $i$th particle, $n_i$ and $n\acute{}_i$, as follows:
\begin{subequations}
\label{eq:index}
\begin{eqnarray}
n_i = \sum_{r_{ij} < 4 a_\mathrm{p}}^{N} 1,
\label{subeq:index1}
\\
n\acute{}_i = \sum_{r_{ij} < 4 a_\mathrm{p}}^{N} n_j,
\label{subeq:index2}
\end{eqnarray}
\end{subequations}
where $r_{ij}$ is the interparticle distance between the $i$th and $j$th particles.
Here, the $n$ of Eq.~(\ref{subeq:index1}) is the number of particles within two particles ($= 4 a_\mathrm{p}$), whereas $n\acute{}$ indicates the number of indirect neighbors through $n$.
In other words, $n\acute{}$ yields the quasi-local particle number density.
According to the value of $n$ or $n\acute{}$, the particle constituting the head is detected, as shown in Figs.~\ref{fig:examples}(a) and (b).
The procedure for $n$ (or, $n\acute{}$) is described below.
The value of $n$ depends on the aggregation size (i.e., head); hence, we set the criterion for judging the constituent particle of the head using the maximum value $n_{\mathrm{max}}$ calculated at each aggregation part.
If the $i$th particle satisfies $n_i / n_{\mathrm{max}} > 1/4$, this particle is assumed to be the constituent of the head.
We apply the same procedure to $n\acute{}$.
Figures~\ref{fig:examples}(a) and (b) show the head defined according to $n$ and $n\acute{}$, respectively.
The blue particle satisfies the criterion above, that is, it is the constituent of the head.
Compared to the $n$ and $n\acute{}$ criteria, the difference is obvious.
The head extends backward along the outside edge of the granular flow in the case of $n$ (Fig.~\ref{fig:examples}(a)), whereas the head is at the front of the granular flow in the case of $n\acute{}$ (Fig.~\ref{fig:examples}(b)).
In this study, we adopt the criterion with $n\acute{}$ as the definition of the head.
To measure the head size, we select four targets from the constituent particles of the head, as shown in Fig.~\ref{fig:examples}(c).
Two green particles are selected as the outmost positions in the $y$ direction, and the others are selected as the forefronts of the outer and inner positions in the $x$ direction.
From these particles, the width and layer thickness of the head, $w_\mathrm{h}$ and $l_\mathrm{h}$, respectively, are measured (Fig.~\ref{fig:examples}(c)).
The temporal variation of the head size and the head size distribution are explained in the next Section~\ref{subsec:characteristics}.

\subsection{Characteristics of head size in $x$--$y$ plane}
\label{subsec:characteristics}

\begin{figure}[t]
\includegraphics[width=1\linewidth]{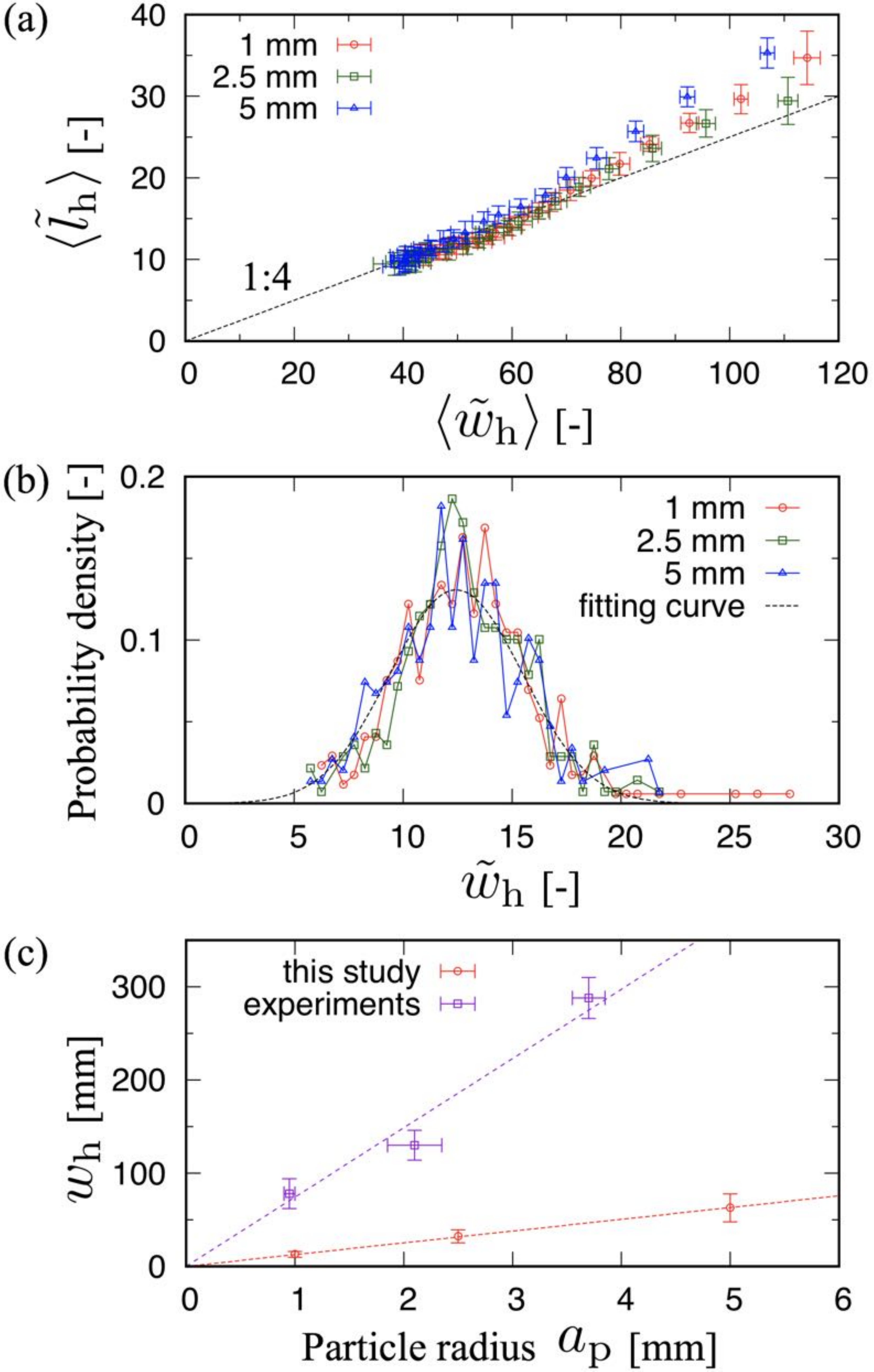}%
\caption{
Variabilities of head size defined in Fig.~\ref{fig:examples}.
Here, $\tilde{w}_\mathrm{h}$ and $\tilde{l}_\mathrm{h}$ are the width and layer thickness of the head normalized by the particle radius $a_\mathrm{p}$.
(a) Time evolutions of $\tilde{w}_\mathrm{h}$ and $\tilde{l}_\mathrm{h}$ in circular setup.
The mean and standard deviation of 20 simulations are denoted by a point and an error bar, respectively.
The data are shown at a constant time interval depending on $a_\mathrm{p}$:
$\Delta t = 10^{-3}$, $5 \times 10^{-4}$, and $2.5 \times 10^{-4}$ s ($a_\mathrm{p} =$ 1, 2.5, and 5 mm).
(b) Probability densities of $\tilde{w}_\mathrm{h}$ in rectangular setup.
Each line is calculated from the final states of 20 simulations.
The dashed black line is a Gaussian distribution fitting the data of $a_\mathrm{p} =$ 5 mm.
(c) Comparison of head width with previous experiments \cite{nohguchi2009vortex}.
The mean and standard deviation of (b) are shown.
The dashed lines are linear functions fitting the simulation and experimental data.
\label{fig:variabilities}}
\end{figure}
Based on the quasi-local particle number density $n\acute{}$ of Eq.~(\ref{subeq:index2}), we study the temporal variation of the head size from the granular flow in the circular shape (see Section~\ref{subsec:circular}).
Because the simulation time is different at each particle radius $a_\mathrm{p}$, the width and layer thickness of the head, $w_\mathrm{h}$ and $l_\mathrm{h}$, respectively, are measured at different time intervals depending on $a_\mathrm{p}$: $\Delta t =$ 1, 0.5, and 0.25 ms ($a_\mathrm{p} =$ 1, 2.5, and 5 mm).
Hence, the number of data is fixed as 20 in each simulation.
Figure~\ref{fig:variabilities}(a) shows the temporal variations of $\tilde{w}_\mathrm{h}$ and $\tilde{l}_\mathrm{h}$ normalized by $a_\mathrm{p}$.
The point and error bar denote the mean and standard deviation of 20 simulations, respectively.
Both $\tilde{w}_\mathrm{h}$ and $\tilde{l}_\mathrm{h}$ decrease with time; the decreasing rate is high in the early stage of the simulations.
This trend is similar to the normalized frontal speed $\tilde{v}_\mathrm{f}$ of the granular flow, as shown in Fig.~\ref{fig:temporal}(a).
Moreover, during the head shrinkage, the ratio between width and layer thickness is maintained ($\tilde{w}_\mathrm{h} / \tilde{l}_\mathrm{h} = 4$) except in the early stage.
We found that the similarity law for the head shape holds in the low-density granular flow.
Next, we verify the head size distribution on the multihead structure of rectangular shape (see Section~\ref{subsec:rectangular}).
In this analysis, the head size is measured at the end of the simulations, where the head elongates downward, as shown in Fig.~\ref{fig:flow}(c).
In Fig.~\ref{fig:flow}(c), the blue particles denote the constituent of the head and multiple heads are detected.
The measurement of layer thickness $l_\mathrm{h}$ is difficult in this situation; hence, only the width $w_\mathrm{h}$ is measured as the head size.
For the number of heads, the mean and standard deviation of 20 simulations at $a_\mathrm{p} =$ 1, 2.5, and 5 mm are $17.20 \pm 1.69$, $13.95 \pm 1.94$, and $14.85 \pm 1.35$, respectively.
Figure~\ref{fig:variabilities}(b) shows the probability density of the normalized width $\tilde{w}_\mathrm{h}$.
Here, $\tilde{w}_\mathrm{h}$ are distributed symmetrically and the distribution shape is independent of $a_\mathrm{p}$.
This distribution is fitted well by the Gaussian distribution indicated by the dashed black line in Fig.~\ref{fig:variabilities}(b).
Although the fluctuation of $\tilde{w}_\mathrm{h}$ might be caused by the inhomogeneity of the initial condition, the flow instability at the moving front of the granular flow exhibits the characteristic wavelength.
For the width of the head $w_\mathrm{h}$, we compare our result qualitatively with the previous experimental result using polystyrene particles~\cite{nohguchi2009vortex}.
In the paper, flow instability resembling a wavy pattern was observed, and the frontal radius was estimated by fitting the circle against each head (see Table~1 of Ref.~\cite{nohguchi2009vortex}).
In this study, we assume that $w_\mathrm{h}$ corresponds to the diameter of the frontal shape (i.e., 2$\times$radius) in previous experiments.
Figure~\ref{fig:variabilities}(c) shows the relationship between particle radius $a_\mathrm{p}$ and $w_\mathrm{h}$ in this study and those of previous experiments.
The point and error bar denote the mean and standard deviation, respectively; particularly, the data in this study are calculated from the distribution as shown in Fig.~\ref{fig:variabilities}(b).
Although the measured $w_\mathrm{h}$ in this study is much smaller than that in the experiments because of the small number of particles in a two-dimensional plane, each data are fitted well by the linear function (dashed line in Fig.~\ref{fig:variabilities}(c)).
We found that $w_\mathrm{h}$ is proportional to $a_\mathrm{p}$ in both cases.

\subsection{Frontal angle in $x$--$z$ plane}
\label{subsec:frontal}

\begin{figure}[t]
\includegraphics[width=1\linewidth]{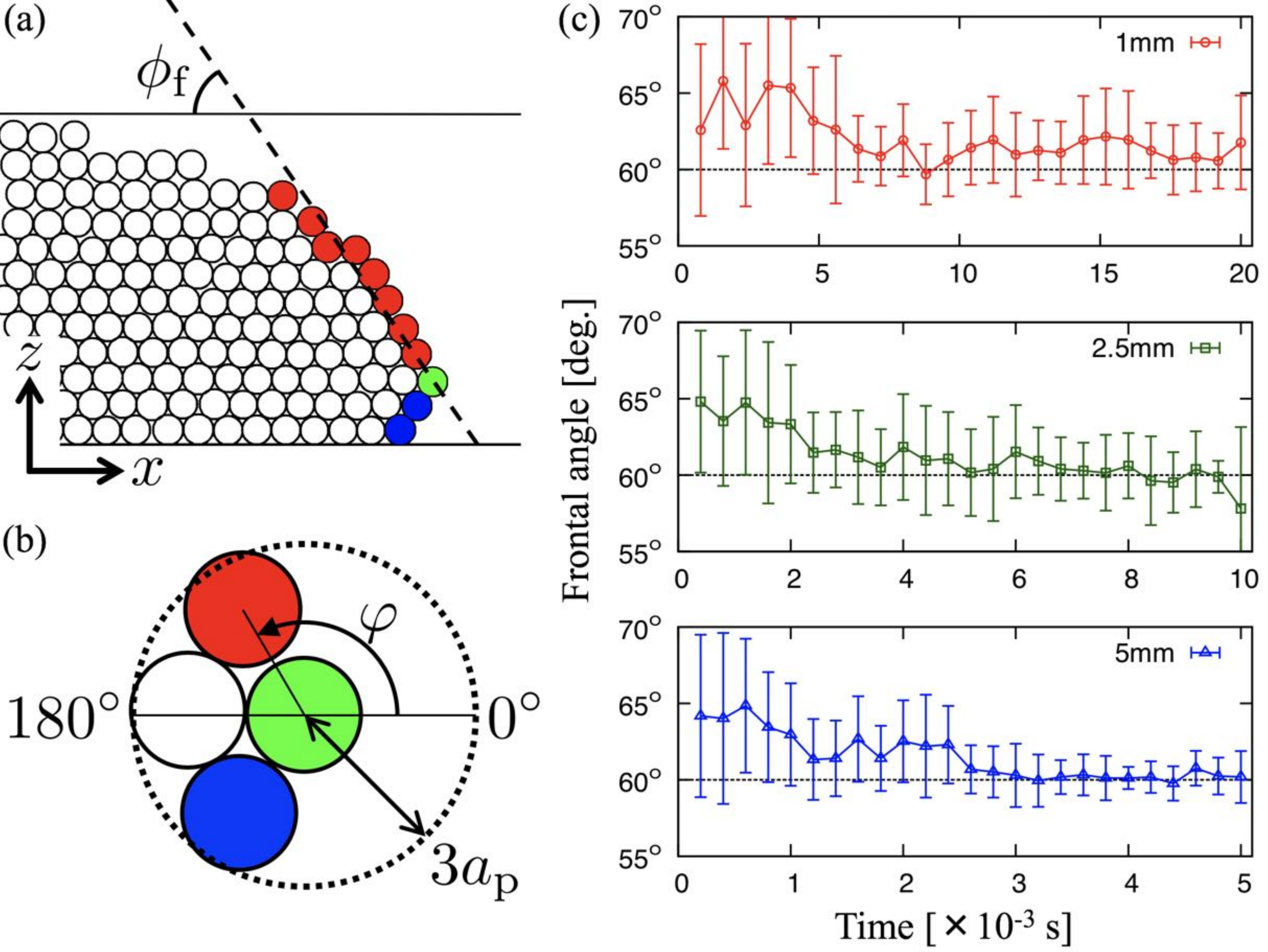}%
\caption{
Definition and temporal variations of frontal angle $\phi_\mathrm{f}$ in $x$--$z$ plane (triangular setup).
(a) and (b)
The green particle is at the forefront of the granular flow, and the neighbor particle is selected within $3 a_\mathrm{p}$ and $\varphi = 0^{\circ}\text{-}180^{\circ}$.
The red particle selected similarly and the green particle contribute to $\phi_\mathrm{f}$, whereas the blue particle is excludable.
The dashed black line is drawn by the least-squares method against the coordinates of the red and green particles.
(c)
The mean and standard deviation of 20 simulations are denoted by a point and error bar, respectively.
\label{fig:definition}}
\end{figure}
As mentioned in Section~\ref{subsec:triangular}, the low-density granular flow is simulated in the $x$--$z$ plane and the head--tail structure is formed.
The shape of the head was observed from the side view of the granular flow in previous experiments with ping-pong balls or polystyrene particles~\cite{mcelwaine2001ping,mcelwaine2005rotational,jackson2017identification}.
Particularly, the frontal angle of the head generally exhibits a high elevation angle to the slope that becomes a constant angle ($60^\circ$) according to the kinematic theory (see Ref.~\cite{mcelwaine2005rotational}).
Hence, we measure the frontal angle of the head from the flow pattern, as shown in Figs.~\ref{fig:formation}(b) and (c), to compare with the fact above.
The definition of frontal angle $\phi_\mathrm{f}$ is based on the coordinate of the particle located at the surface of the head.
Figure~\ref{fig:definition}(a) shows the typical shape of the head and the corresponding $\phi_\mathrm{f}$.
As the procedure to measure $\phi_\mathrm{f}$, we first select the forefront particle of the granular flow in the $x$ direction: green particle in Fig.~\ref{fig:definition}(a).
Next, we search the constituent of the head surface around the green particle, as shown in Fig.~\ref{fig:definition}(b), where the search range is three particle radiuses and the search angle against the slope is $\varphi = 0^\circ\text{-}180^\circ$.
By repeating this search, the red particles in Fig.~\ref{fig:definition}(a) are detected.
A fitting line is obtained from the coordinates of the green and red particles using the least-squares method; subsequently, $\phi_\mathrm{f}$ is estimated.
It is noteworthy that the blue particles below the green particle are excluded from the estimation.
Figure~\ref{fig:definition}(c) shows the temporal variation of the frontal angle $\phi_\mathrm{f}$ at each particle radius $a_\mathrm{p}$ in the triangular shape.
The mean and standard deviation of 20 simulations are denoted by the point and error bar, respectively.
In the initial condition as shown in Fig.~\ref{fig:formation}(a), $\phi_\mathrm{f}$ is estimated as approximately $90^\circ$.
Subsequently, $\phi_\mathrm{f}$ approaches $60^\circ$ from the larger angle as time progresses, and the standard deviation of $\phi_\mathrm{p}$ decreases with time in all cases.
This convergence angle of $60^\circ$ is consistent with the frontal angle reported in previous studies~\cite{mcelwaine2005rotational}.
However, $\phi_\mathrm{f} = 60^\circ$ in this study resulted in the crystallization of particles in the two-dimensional simulation (Figs.~\ref{fig:formation}(b) and (c)), although we used particles with a polydispersity of $\pm5~\%$ to avoid crystallization.
Even though a three-dimensional simulation was performed, such a crystallization is to be expected.

\subsection{Nondimensionalization for model}
\label{subsec:nondimensionalization}

Finally, we discuss the dimensionless parameter to determine the dynamics of the low-density granular flow in our model.
In the discussions thus far, the similar flow pattern can be found independent of the particle radius $a_\mathrm{p}$ (see Section~\ref{sec:results}).
Additionally, the head size is well normalized by $a_\mathrm{p}$, as shown in Figs.~\ref{fig:variabilities}(a) and (b), and the frontal speed is normalized by the single-particle velocity $v_\mathrm{s}$ depending on $a_\mathrm{p}$, as shown in Figs.~\ref{fig:temporal}(a), \ref{fig:flow}(d), and \ref{fig:formation}(d).
Therefore, we perform the nondimensionalization of the governing equations in Eq.~(\ref{eq:governing}).
We search the characteristic length and speed, $\alpha$ and $\beta$, respectively, from Eqs.~(\ref{eq:gravity})--(\ref{eq:governing}), as follows:
\begin{eqnarray}
\tilde{\bm{r}} = \frac{\bm{r}}{\alpha},~
\tilde{\bm{v}} = \frac{\bm{v}}{\beta},~
\tilde{t} = \frac{t}{\alpha / \beta},
\label{eq:characteristic}
\end{eqnarray}
where $\bm{r}$, $\bm{v}$, and $t$ are the particle coordinates, particle velocity, and time, respectively.
Here, we use the particle radius as the characteristic length: $\alpha = a_\mathrm{p}$, and we consider the monodisperse system hereinafter.
Equation~(\ref{subeq:governing1}) that describes the velocity of the $i$th particle $\bm{v}_i$ is rewritten as follows:
\begin{eqnarray}
\tilde{\bm{v}}_i
&=&
\frac{1}{6 \pi} \frac{\bm{F}_i^{g+r}}{\mu a_\mathrm{p} \beta}
+ \sum_{j \ne i}^N \frac{\bm{u}_i(j)}{\beta}
\nonumber
\\
&=&
\frac{\tilde{\bm{F}}_i^{g+r}}{6 \pi}
+ \sum_{j \ne i}^N \tilde{\bm{u}}_i(j),
\label{eq:dimensionless1}
\end{eqnarray}
where $\tilde{\bm{F}}_i^{g + r} = \tilde{\bm{F}}_i^{g} + \tilde{\bm{F}}_i^{r}$ is the sum of the dimensionless gravity and dimensionless repulsive force of the $i$th particle, and $\tilde{\bm{u}}_i (j)$ is the dimensionless induced velocity generated by the $j$th particle at the coordinate of the $i$th particle.
Furthermore, $\tilde{\bm{u}}_i (j)$ is expressed using Eqs.~(\ref{subeq:drag2}) and (\ref{subeq:drag3}) as follows:
\begin{eqnarray}
\tilde{\bm{u}}_i(j)
&=&
\frac{1}{\beta} \frac{1}{8 \pi \mu} \bm{\mathrm{J}}(\bm{r}) \cdot \bm{F}_j^{g+r}
\nonumber
\\
&=&
\frac{1}{8 \pi} \bm{\mathrm{J}}(\tilde{\bm{r}}) \cdot \tilde{\bm{F}}_j^{g+r}.
\label{eq:dimensionless2}
\end{eqnarray}
In this equation, because the dimension of the Rotne--Prager tensor $\bm{\mathrm{J}}(\bm{r})$ is the inverse of $\bm{r}$, we obtain an equality: $\bm{\mathrm{J}}(\bm{r}) = \bm{\mathrm{J}}(\tilde{\bm{r}}) / a_\mathrm{p}$.
The dimensionless parameter is gathered into $\tilde{\bm{F}}_i^{g + r}$ in Eqs.~(\ref{eq:dimensionless1}) and (\ref{eq:dimensionless2}).
The dimensionless repulsive force $\tilde{\bm{F}}_i^r$ is expressed using Eq.~(\ref{eq:repulsive}) as follows:
\begin{eqnarray}
\tilde{\bm{F}}_i^r
&=&
- \frac{1}{\mu a_\mathrm{p} \beta} k_n \delta_{ij} \bm{n}_{ij}
\nonumber
\\
&=&
- \tilde{\delta}_{ij} \bm{n}_{ij}
~~~~(\beta = k_n / \mu),
\label{eq:dimensionless3}
\end{eqnarray}
where $\tilde{\delta}_{ij} = \delta_{ij} / a_\mathrm{p}$ is the dimensionless overlap between the $i$th and $j$th particles.
To reduce the number of variables, the characteristic speed is given as $\beta = k_n / \mu$.
Using the specific form of $\beta$ and Eq.~(\ref{eq:gravity}), the dimensionless gravity $\tilde{\bm{F}}_i^g$ is rewritten as follows:
\begin{eqnarray}
\tilde{\bm{F}}_i^g
&=&
- \frac{1}{\mu a_\mathrm{p} \beta} V_i g (\rho_\mathrm{g} - \rho_\mathrm{f}) e_\theta
\nonumber
\\
&=&
- \gamma \tilde{V}_i e_\theta
~~~~\left(\gamma = \frac{g (\rho_\mathrm{g} - \rho_\mathrm{f}) a_\mathrm{p}^2}{k_n}\right),
\label{eq:dimensionless4}
\end{eqnarray}
where $\tilde{V}_i = V_i / a_\mathrm{p}^3$ is the dimensionless volume of the $i$th particle and $\gamma$ is the dimensionless parameter in our model.
From the discussions above, we discovered that our model can be simplified to dimensionless equations with one dimensionless parameter $\gamma$.
The physical meaning of $\gamma$ is the ratio of gravity divided by the excluded volume effect between particles, that is, $\bm{F}^g / \bm{F}^r$ determines the dynamics of the low-density granular flow in our model.
By substituting the values used for the simulations in $\gamma$, we obtain $\gamma=1.8 \times 10^{-5}$, $1.2 \times 10^{-4}$, and $4.6 \times 10^{-4}$ ($a_\mathrm{p} =$ 1, 2.5, and 5 mm).
Therefore, the numerical data in this study can be explained in terms of $\gamma \ll 1$ (i.e., $\bm{F}^g \ll \bm{F}^r$).


\section{Conclusion}
\label{sec:conclusion}

In this study, three types of pattern formations for a low-density granular flow were reproduced qualitatively using the single particle model: crescent-shaped head, wavy pattern due to flow instability, and head--tail structure.
This model considered the gravity, repulsive force, and drag force due to the fluid as forces acting on the particle, which was the minimum number of factors to simplify the natural phenomenon with complicated behaviors.
The frontal speed of the head decreased with time, and was related positively to the head shrinkage (or number of constituent particles).
Moreover, granular vortex convection was formed at the head in the $x$--$y$ plane parallel to the slope, whereas the particle moved from the upper and rear parts of the head to the tail in the $x$--$z$ plane perpendicular to the slope.
According to the quasi-local particle number density, the width and layer thickness of the head were defined as the characteristic head size.
The ratio (width/layer thickness) was maintained ($\approx$ 4) while the head shrunk, and the size distribution of the width obeyed the Gaussian distribution.
For a comparison with previous experiments regarding the low-density granular flow, we found a linear relationship between the head width and particle radius; additionally, the frontal angle of the head converged on $60^\circ{}$.
Finally, these results were determined by one dimensionless parameter (the ratio of gravity to repulsive force) through the nondimensionalization performed for the model.
This model is not realistic yet; however, we believe we pioneered a particle-based approach to elucidate the dynamics of low-density granular flow in this study.


\begin{acknowledgments}
This work was supported by JPSJ KAKENHI grant numbers 11J07296, 14J03528, 17K14353.
We would like to thank Editage (www.editage.jp) for English language editing.
\end{acknowledgments}

\bibliography{Simple_niiya}

\end{document}